\documentclass[letterpaper]{article} 
\usepackage[preprint]{aaai2027}  
\usepackage[hyphens]{url}  
\usepackage{graphicx} 
\usepackage{amsmath}
\usepackage{amssymb}
\usepackage{booktabs}
\usepackage{array}
\usepackage{pdfpages}
\usepackage{natbib}  
\usepackage{caption} 
\makeatletter
\patchcmd{\@maketitle}
  {Corresponding author\ifaaai@corrmulti{s}\fi.}
  {Corresponding authors.}
  {}{}
\makeatother

\newcommand{\paperauthors}{Yujian Ma\textsuperscript{1}, Jinqiu Sang\textsuperscript{2}\corresponding, Ruizhe Li\textsuperscript{3}\corresponding, Jiaao Yu\textsuperscript{2}, Ang Li\textsuperscript{4}}
\newcommand{\paperaffiliations}{%
\textsuperscript{1}Shanghai Institute of Artificial Intelligence for Education, East China Normal University, China\\
\textsuperscript{2}School of Computer Science and Technology, East China Normal University, China\\
\textsuperscript{3}School of Computer Science, University of Birmingham, UK\\
\textsuperscript{4}School of Information Engineering, Chang'an University, China%
}

\title{From Semantics to Readout: Mechanistic Understanding of Audio Tokens after Fine-Tuning for Temporal Audio Grounding}
\author{\paperauthors}
\affiliations{\paperaffiliations}

\begin{document}

\maketitle

\begin{abstract} 
    Large audio-language models (LALMs) convey acoustic evidence to language decoders through native audio tokens, yet the internal roles of these tokens remain poorly understood. Using temporal audio grounding as a diagnostic setting, we examine how language-model fine-tuning affects the layerwise semantics, decoder accessibility, and temporal output alignment of native audio-token states through four complementary analyses: query-conditioned token semantics, calibrated token readout, temporal-window probes, and residual-delta erasure during generation. Alongside substantial improvements in temporal localization, semantic analysis of Qwen2.5-Omni shows that latent evidence for queried events is already present before fine-tuning and that the audio tokens most strongly aligned with the queried event appear at similar temporal positions before and after fine-tuning. After fine-tuning, event-related information in audio tokens becomes more accessible to the decoder, especially in early and middle layers, and a cross-checkpoint control shows that this improvement arises primarily from decoder adaptation. Temporal probes show that the base checkpoint already contains recoverable information about annotated windows and that fine-tuning mainly improves alignment with each checkpoint's own predicted temporal support. Residual-delta erasure further shows that removing audio-token updates within predicted windows harms timestamp generation more than removing the same number of randomly selected updates. The same broad improvements in decoder readability and prediction alignment also appear in Qwen2-Audio. Together, these results support a semantics-to-readout account in which grounding fine-tuning helps the decoder read existing event evidence and connect it more reliably to temporal outputs. 
\end{abstract}

\section{Introduction}

Large audio-language models (LALMs) connect acoustic signals with the reasoning and generation capabilities of large language models \cite{tang2024salmonn,chu2024qwen2audio,kong2024audioflamingo,wang2024audiobench}. They convert continuous waveforms into audio tokens, the interface through which acoustic evidence enters language reasoning. Yet these native audio tokens remain poorly understood: they support downstream behavior, but what they encode, when this information becomes readable, and how it reaches final text remain unclear.

Temporal grounding gives a controlled way to probe this question. General audio understanding asks what event occurs; temporal grounding also asks when. This makes task fine-tuning an informative intervention because the model must connect event semantics to explicit temporal outputs. Since general LALMs are not primarily trained for precise timestamp grounding \cite{tang2024salmonn,chu2024qwen2audio}, recent evaluations report persistent failures in localization, duration estimation, temporal reasoning, and timestamp prediction \cite{ahia2025blab,kulkarni2026closer,yao2025notinsync}.

\begin{figure*}[!t]  
\centering
\includegraphics[width=\textwidth]{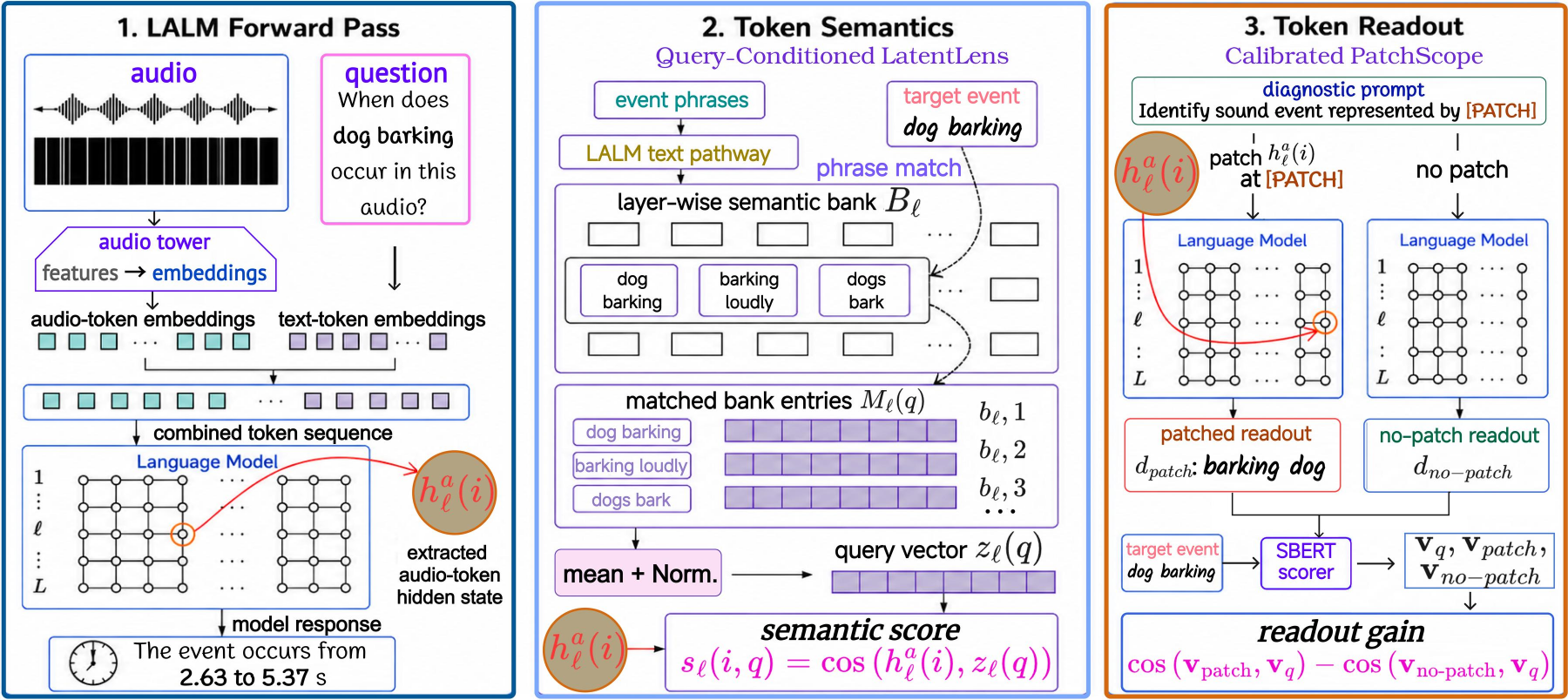}  

\caption{
Overview of the core audio-token semantics-to-readout analysis pipeline.
(1) \textit{LALM forward pass:} the audio clip and question \(q\) form a mixed audio-text sequence, from which we extract layerwise audio-token states \(h_\ell^a(i)\).
(2) \textit{Token semantics:} task event phrases are encoded by the LALM text pathway to build \(\mathcal{B}_\ell\) with text-side states \(b_{\ell,j}\). Phrase matching selects \(\mathcal{M}_\ell(q)\) for the target event phrase \(e(q)\); mean normalization gives \(z_\ell(q)\), and \(s_\ell(i,q)\) scores each audio token.
(3) \textit{Token readout:} \(h_\ell^a(i)\) is inserted at the diagnostic \texttt{[PATCH]} site and decoded with a no-patch baseline. Sentence-BERT (SBERT) embeds the target event phrase from \(q\) and both readouts as \(\mathbf{v}_{q}\), \(\mathbf{v}_{\mathrm{patch}}\), and \(\mathbf{v}_{\mathrm{no\text{-}patch}}\); their cosine difference gives calibrated readout gain.
}
\label{fig:timeground-overview}
\end{figure*}

Recent work improves temporal grounding by making time explicit through time-aware audio representations, audio prompts, timestamp-interleaved training sequences, or grounding objectives \cite{wang2025timeaudio,shi2026towards,sun2026spotsound,ghosh2026framelevel}. These methods show that stronger temporal supervision and interfaces help. However, they mainly optimize timestamp interfaces and outputs, often using fine-tuned LALMs as baselines before adding explicit timestamp structure. This leaves a key internal question underexplored: how does task fine-tuning change the semantics and decoder readout of native audio-token representations?

As illustrated in Figure~\ref{fig:timeground-overview}, we use fine-tuning for temporal grounding to compare native audio-token states before and after task adaptation. Rather than treating fine-tuning as a black box that simply adds temporal capabilities, we organize the analysis around four linked research questions: \textbf{RQ1}: What query-relevant event evidence is already present in native audio-token states before grounding fine-tuning, and how stable are its temporal locations across checkpoints? \textbf{RQ2}: How does fine-tuning affect the decoder's ability to read this evidence, and do the resulting gains arise primarily from intermediate-state changes or downstream decoder adaptation? \textbf{RQ3}: How do audio-token states correspond to annotated event windows and to each checkpoint's own predicted temporal support? \textbf{RQ4}: Do residual updates within predicted windows affect timestamp generation? Together, these questions trace the progression from event semantics to decoder readout, temporal output alignment, and generation-time relevance.

The main contributions are summarized as follows:
\begin{itemize}
\item \textbf{Controlled diagnostic setting.} We use temporal grounding fine-tuning to examine changes in native audio-token states beyond timestamp accuracy.
\item \textbf{Semantics-to-readout framework.} We combine query-conditioned token semantics, calibrated token readout, temporal-window probes, and residual-delta erasure to distinguish event representation, decoder accessibility, checkpoint-specific output consistency, and functional relevance in generation.
\item \textbf{Empirical answers.} \textbf{RQ1}: Base checkpoints already contain latent event evidence at largely stable temporal locations. \textbf{RQ2}: Fine-tuning improves decoder readability primarily through downstream decoder adaptation. \textbf{RQ3}: It strengthens consistency with predicted temporal support more than recoverability of annotated windows. \textbf{RQ4}: Predicted-window residual updates are functionally relevant to timestamp generation.
\end{itemize}

\section{Related Work}

\paragraph{Temporal Audio Grounding}
Temporal audio grounding localizes the time intervals of queried sound events. It is closely related to sound event detection, which localizes predefined sound classes from strong or weak temporal labels \cite{bhosale2024diffsed,cornell2024dcase}. Text query grounding generalizes this setting to natural-language event phrases, as in AudioGrounding's phrase-level links among clips, captions, event phrases, and timestamp intervals \cite{audiogrounding_dataset}. Because it conditions on phrases, this formulation naturally fits LALMs, which jointly condition on audio and text and generate textual responses.

For LALMs, grounding methods typically improve timestamp prediction by making time explicit in the interface or objective. TimeAudio and SpotSound use temporal markers, time encodings, timestamp sequences, or specialized temporal supervision \cite{wang2025timeaudio,sun2026spotsound}; TimePro-RL injects timestamp embeddings and optimizes temporal alignment with rewards \cite{shi2026towards}; and tool use at the frame level exposes finer temporal operations for audio-language models \cite{ghosh2026framelevel}. These works mainly study how temporal coordinates should be represented, supervised, or generated. In contrast, we use temporal grounding as a diagnostic setting for studying how task fine-tuning changes native audio-token semantics and decoder readout.

\paragraph{Audio Tokens as Interfaces}
LALMs follow a common multimodal design in which audio inputs are encoded, projected into the hidden space of the language model, and placed in the decoder context as token representations. Acoustic encoders and projection modules produce native audio-token states that condition text generation, enabling audio question answering, captioning, dialogue, and general audio understanding \cite{tang2024salmonn,chu2024qwen2audio,kong2024audioflamingo,qwen25omni}. These states form the internal interface through which acoustic evidence reaches the language decoder. Benchmarks such as AudioBench and AIR-Bench evaluate external capabilities across speech, environmental sound, music, and open-ended audio interaction \cite{wang2024audiobench,yang2024airbench}. They do not, however, explain how fine-tuning for a task changes what event evidence native audio-token states organize, or whether that evidence becomes more accessible to the language decoder.

\paragraph{Token Evidence and Readout}
Methods for interpreting hidden states provide distinct ways to inspect internal evidence and its accessibility. Linear probing tests whether information can be recovered, although recoverability alone does not show that the evidence is used during generation \cite{hewitt2019control}. LatentLens compares hidden states with contextualized semantic references \cite{krojer2026latentlens}, while Patchscopes use diagnostic prompts to verbalize hidden states \cite{ghandeharioun2024patchscopes}. Causal tracing and activation-patching interventions test whether hidden-state components affect model outputs \cite{meng2022locating}. These methods separate recoverability, semantic alignment in representation space, decoder readout, and functional relevance.

Interpretability work in audio has begun to examine internal representations beyond final task accuracy. AudioLens traces auditory attribute evidence in LALMs through projections into vocabulary space \cite{yang2025audiolens}; Beyond Transcription uses probing and interventions to study automatic speech recognition representations \cite{glazer2026beyond}; Ma et al. analyze how LoRA adaptation reshapes Whisper encoder representations for speech emotion recognition through probing, logit-lens, and representation-similarity analyses \cite{ma2026behind}; causal tracing examines audio-text fusion across layers and token positions \cite{chen2026causal}; and audio-visual stress tests show that multimodal LLMs may fail to surface auditory evidence in generated answers even when sound information is present \cite{selvakumar2026audio}. These studies reveal semantic evidence in audio representations and possible gaps between audio information and generated behavior, but they do not directly examine how task fine-tuning changes native audio-token semantics, decoder readout, and alignment with explicit temporal outputs. We address this gap by treating temporal-grounding fine-tuning as a controlled intervention and comparing representation, readout, and output correspondence across checkpoints.

\section{Method}
\label{sec:method}

Our diagnostics compare native audio-token states before and after grounding fine-tuning using query-conditioned semantics, calibrated readout, output-facing temporal probes, and an intervention during generation. These diagnostics characterize representational alignment, decoder accessibility, output consistency, and functional relevance for timestamp generation.

\subsection{Audio-Token Interface}
\label{sec:method_prelim}

Given an audio clip $x$ and a natural-language audio question, we denote the
full question by \(q\) and the target event phrase extracted from it by \(e(q)\).
The extracted event phrase is used for matching and scoring at the event level.
Our analysis focuses on the internal
audio-token representations that connect acoustic evidence to textual responses
about events, rather than on the final generated answer alone.

The processor constructs a mixed audio-text input sequence containing text
tokens from the question and, for the waveform, audio placeholder tokens specific
to the model. The audio tower encodes the waveform into audio features, which are
projected into representations associated with the corresponding audio-token
positions. Let $\mathcal{A}$ denote the native audio-token positions in this
mixed sequence, identified by the audio-token id used by the model. For the $i$-th
audio token and layer $\ell$, we denote its hidden state as $h_\ell^a(i)$, where
$i$ indexes audio-token positions and $\ell$ indexes LALM layers.

These layerwise audio-token states are the focus of our analysis.

\subsection{Query-Conditioned Token Semantics}
\label{sec:method_latentlens}

We adapt contextualized semantic reference retrieval to analyze audio-token
semantics conditioned on the question. The representations under inspection are
native audio-token hidden states $h_\ell^a(i)$, and the semantic reference is
the target sound event specified by \(e(q)\) rather than an
unconstrained text neighborhood.

For each layer $\ell$, we construct an event-phrase semantic bank
$\mathcal{B}_\ell=\{(c_j,b_{\ell,j})\}$ from normalized event expressions in
AudioGrounding-QA. Each row \(j\) stores the contextual state \(b_{\ell,j}\) of
one text token from the LALM text pathway and its source expression \(c_j\).
The term ``event phrase'' also covers simple event clauses. Given the event
expression \(e(q)\), we
retrieve a matched subset $\mathcal{M}_\ell(q)\subseteq\mathcal{B}_\ell$ using
the \texttt{exact}, \texttt{contains}, and \texttt{token\_overlap\_k} matching
cascade. The semantic vector conditioned on the question is the normalized mean of the matched bank
embeddings:
\[
z_\ell(q)
=
\mathrm{norm}\!\left(
\frac{1}{|\mathcal{M}_\ell(q)|}
\sum_{(c_j,b_{\ell,j})\in \mathcal{M}_\ell(q)}
b_{\ell,j}
\right).
\]

We score each audio-token state by its cosine similarity to this semantic
vector, which we call the target event alignment score:
\[
s_\ell(i,q)
=
\cos\!\left(h_\ell^a(i), z_\ell(q)\right).
\]
High values indicate that the audio-token state is close to contextualized text
representations of the queried event. The diagnostic is checkpoint-relative
because both the audio-token states and text-side event references for \(e(q)\) are extracted
from the corresponding checkpoint; it therefore measures within-checkpoint
audio-text alignment rather than isolating changes in audio states alone. We define the top-scoring set
$\mathcal{H}_\ell(q)$ as the top \(M=20\) audio tokens under this score. In
comparisons with predicted windows, this statistic measures how strongly the
event tokens with the highest scores align with the model's generated temporal support.
Thus, the audio-token semantics diagnostic provides a target event alignment
score \(s_\ell(i,q)\) for comparing token-level evidence against annotated and
predicted temporal windows.
Appendix~B further reports semantic bank validation and examples of the
\texttt{exact}, \texttt{contains}, and \texttt{token\_overlap\_k} matching
cascade used to construct \(\mathcal{M}_\ell(q)\).

\subsection{Calibrated Token Readout}
\label{sec:method_patchscope}

The audio-token semantics diagnostic measures alignment in representation space,
but it does not test whether the language decoder can express the information
carried by an audio-token state. We therefore use a calibrated hidden-state
insertion diagnostic for audio-token readout: a source audio-token state
$h_\ell^a(i)$ is inserted into a sound event prompt to elicit a
natural language readout.

For each audio-token state $h_\ell^a(i)$, we use a diagnostic text prompt \(p\) to ask the model to identify the sound event represented by an internal audio representation. The prompt contains a placeholder position \(r\), used as the readout site, with hidden state \(h_\ell^p(r)\). At layer \(\ell\), we replace this placeholder state with \(h_\ell^a(i)\), continue the remaining decoder computation, and decode a short event phrase \(d_{\ell,i}^{\mathrm{patch}}\) deterministically.

A diagnostic prompt can itself induce generic or non-informative outputs. To isolate the semantic contribution of the patched audio-token state, we decode the same prompt without any patch, yielding \(d^{\mathrm{no\text{-}patch}}\). We embed the patched readout, no-patch readout, and the target event phrase \(e(q)\) with SBERT~\cite{sentence-bert}, yielding \(\mathbf{v}_{\ell,i}^{\mathrm{patch}}\),
\(\mathbf{v}^{\mathrm{no\text{-}patch}}\), and \(\mathbf{v}_{q}\), respectively.
We define the calibrated semantic readout gain as
\[
\Delta_{\mathrm{readout}}(\ell,i)
=
\cos\!\left(\mathbf{v}_{\ell,i}^{\mathrm{patch}}, \mathbf{v}_{q}\right)
-
\cos\!\left(\mathbf{v}^{\mathrm{no\text{-}patch}}, \mathbf{v}_{q}\right).
\]
Positive \(\Delta_{\mathrm{readout}}\) indicates that patching the audio-token
state improves semantic readout of the target event beyond the no-patch diagnostic
baseline.

This calibrated variant differs from direct output inspection: it asks whether a
specific audio-token hidden state can be expressed by the language decoder as
text semantically matching the queried event. Together with the semantics
diagnostic, it separates semantics in representation space from token readout by
the decoder.

\subsection{Output-Facing Temporal Diagnostics}
\label{sec:method_probe}
\paragraph{Temporal-window probes.}
We use temporal-window probes as standard diagnostic linear classifiers over
frozen audio-token states. For each checkpoint, layer, and label source, hidden
vectors are standardized from the training split and an L2-regularized logistic
probe is trained to predict whether an audio token belongs to a temporal
support region. For an audio token \(i\) at layer \(\ell\), the probe predicts
\[
p_\ell(y_i=1 \mid h_\ell^a(i))
=
\sigma\!\left(w_\ell^\top h_\ell^a(i)+b_\ell\right),
\]
where \(y_i\) indicates whether the token overlaps a target temporal window. We
use two label sources:
\(y_i^{\mathrm{gt}}=\mathbf{1}[\omega_i\cap W^\star\neq\emptyset]\) for
annotated event windows, and
\(y_i^{\mathrm{pred}}=\mathbf{1}[\omega_i\cap \widehat{W}\neq\emptyset]\) for
the model's generated temporal windows, where \(\omega_i\) is the approximate
time span assigned to audio token \(i\). For predicted-window labels,
\(\widehat{W}\) is constructed separately from each checkpoint's parsed
timestamp output, so these probes measure checkpoint-specific consistency with
generated temporal behavior rather than performance on a shared prediction
target.

Layerwise figures report held-out token-level AUROC across all layers as the primary threshold-free metric. Appendix~E reports implementation details, including label ratios, AUPRC, validation-selected F1 and balanced accuracy, and a position-only control. These probes are audio-state-only and do not take the event phrase or query-conditioned vector as input. Thus, annotated-window probes test linear decodability of annotated temporal membership, while predicted-window probes test whether audio-token states are linearly aligned with each checkpoint's own temporal outputs.

\paragraph{Residual-delta erasure.}
To test functional relevance in the original generation context, we apply residual-delta erasure to the fine-tuned Qwen2.5-Omni checkpoint. At a selected layer, we replace the layer output at audio-token positions overlapping the checkpoint's predicted temporal support with the corresponding layer input, thereby removing only that layer's residual update. We compare these predicted-window positions with a size-matched random audio-token control, sampled outside the predicted support when possible. We report the degradation rate, defined as the percentage of instances whose timestamp IoU drops by more than 0.05 relative to unmodified generation; sampling and evaluation details are provided in Appendix~E.


\section{Experiments}
\subsection{AudioGrounding-QA}

We construct AudioGrounding-QA by converting phrase-level temporal annotations from the public AudioGrounding corpus, released with TextToAudioGrounding~\cite{audiogrounding_dataset}, into an audio question-answering format. AudioGrounding is built on AudioCaps/AudioSet clips~\cite{audiocaps,audioset}; each phrase item with timestamped segments becomes one event-centered temporal question and an answer verbalized from the annotated intervals.
After QA construction and cleaning, the dataset contains 9,542/1,052/992 QA instances for train/val/test, respectively (11,586 in total), following the official AudioGrounding split protocol. Table~\ref{tab:data_stats_split} reports split statistics. Here, QA/Aud. measures the number of event-centered questions per audio clip, Avg.\#Seg reports temporal fragmentation through the average number of annotated intervals, and Avg. Cov. (s) denotes the average total duration covered by the queried event on the approximately 10-second timeline. Appendix~A gives construction prompts and QA examples.

\begin{figure*}[!t]
\centering
\makebox[\textwidth][c]{\includegraphics[width=1.04\textwidth]{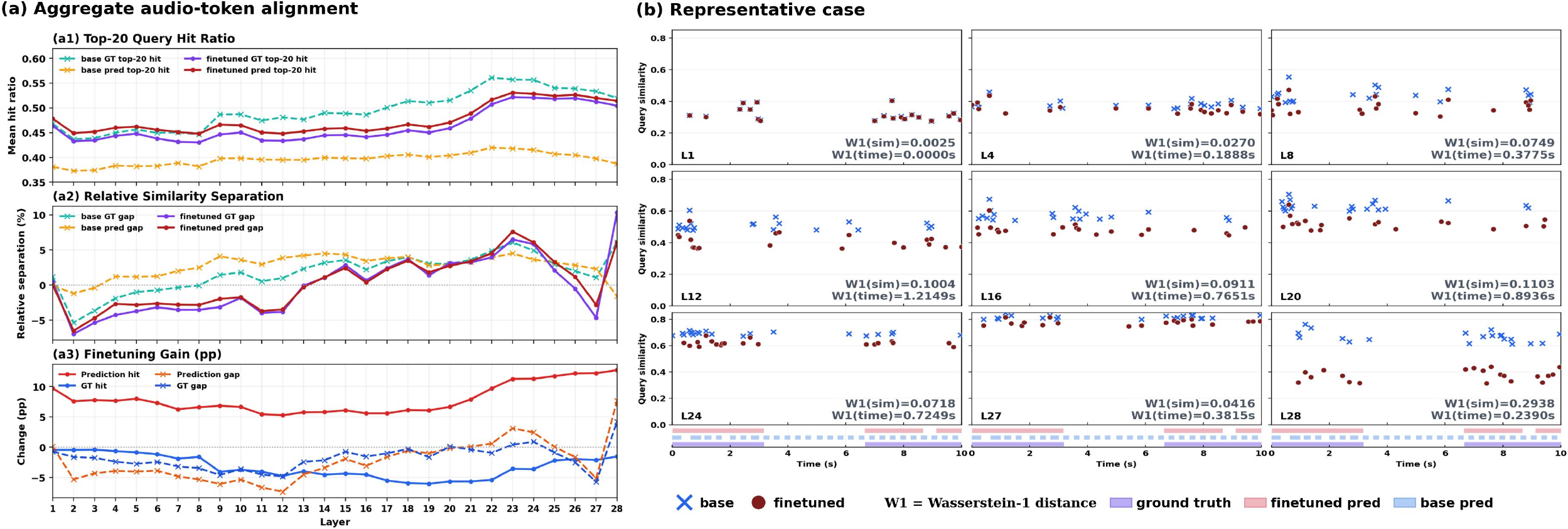}}
\caption{Audio-token semantics analysis.
(a1) Top-20 overlap with annotated and predicted temporal support.
(a2) Inside--outside support score separation.
(a3) Layerwise fine-tuning gains.
(b) A representative Top-20 token case; $W_1$ quantifies cross-checkpoint differences in token time and semantic similarity.}
\label{fig:latentlens}
\end{figure*}

\begin{table}[!t]
\centering
{\small
\setlength{\tabcolsep}{2.3pt}
\begin{tabular}{lccccc}
\toprule
Split & \#QA & \#Aud. & QA/Aud. & Avg.\#Seg & Avg. Cov. (s) \\
\midrule
Train & 9,542 & 3,686 & 2.59 & 1.65 & 5.07 \\
Val   & 1,052 &   483 & 2.18 & 1.77 & 4.95 \\
Test  &   992 &   483 & 2.05 & 1.83 & 4.75 \\
\midrule
Total & 11,586 & 4,652 & 2.49 & 1.67 & 5.01 \\
\bottomrule
\end{tabular}
}
\caption{Statistics of the AudioGrounding-QA dataset.}
\label{tab:data_stats_split}
\end{table}

\subsection{Models and Evaluation}

We fine-tune two LALMs, Qwen2.5-Omni-7B \cite{qwen25omni} and Qwen2-Audio-7B-Instruct \cite{chu2024qwen2audio}, on the AudioGrounding-QA training split using Low-Rank Adaptation (LoRA) \cite{lora}. Qwen2.5-Omni is the primary model for detailed audio-token analysis, and Qwen2-Audio tests whether the pattern replicates across architectures. In both models, LoRA adapters are applied only inside the language model transformer blocks; the audio tower and multimodal audio to language projector remain frozen. Thus, acoustic encoding is fixed while the transformation of audio-token states inside the language model can change. Full fine-tuning and hardware details are provided in Appendix~D, and Appendix~G reports the Qwen2-Audio replication.

All main results are reported on the 992-instance test split. For temporal grounding evaluation, model outputs are parsed into timestamp intervals and compared with annotated event intervals using mean intersection over union (mIoU), F1 at the time level, and recall at IoU thresholds R@0.5, R@0.7, and R@0.9. For the main token analysis on Qwen2.5-Omni, we inspect all 28 Thinker layers using the diagnostics defined in Section~\ref{sec:method}. Appendix~C specifies the readout diagnostic prompt and representative readout examples, Appendix~D provides generation, parsing, and metric details, and Appendix~E gives probe and residual-erasure implementation details for reproducibility.

\subsection{Grounding Performance}

Table~\ref{tab:grounding_perf} shows that fine-tuning substantially improves temporal grounding on AudioGrounding-QA. Qwen2.5-Omni improves from 0.3707 to 0.6817 mIoU and from 0.4416 to 0.7626 F1, with R@0.7 rising from 0.2349 to 0.5817. Qwen2-Audio shows the same trend, improving from 0.3653 to 0.6199 mIoU and from 0.4667 to 0.7195 F1. These gains are consistent across architectures.

\begin{table}[!t]
\centering
{\small
\setlength{\tabcolsep}{2.1pt}
\begin{tabular}{lccccc}
\toprule
Model & mIoU & F1 & R@0.5 & R@0.7 & R@0.9 \\
\midrule
Qwen2.5-Omni base & 0.3707 & 0.4416 & 0.3377 & 0.2349 & 0.1542 \\
Qwen2.5-Omni FT & 0.6817 & 0.7626 & 0.7399 & 0.5817 & 0.3286 \\
Qwen2-Audio base & 0.3653 & 0.4667 & 0.2873 & 0.1704 & 0.0927 \\
Qwen2-Audio FT & 0.6199 & 0.7195 & 0.6714 & 0.4556 & 0.2157 \\
\bottomrule
\end{tabular}
}
\caption{Temporal grounding on AudioGrounding-QA. FT: fine-tuned checkpoint.}
\label{tab:grounding_perf}
\end{table}

\begin{figure*}[!t]
\centering
\makebox[\textwidth][c]{\includegraphics[width=1.04\textwidth]{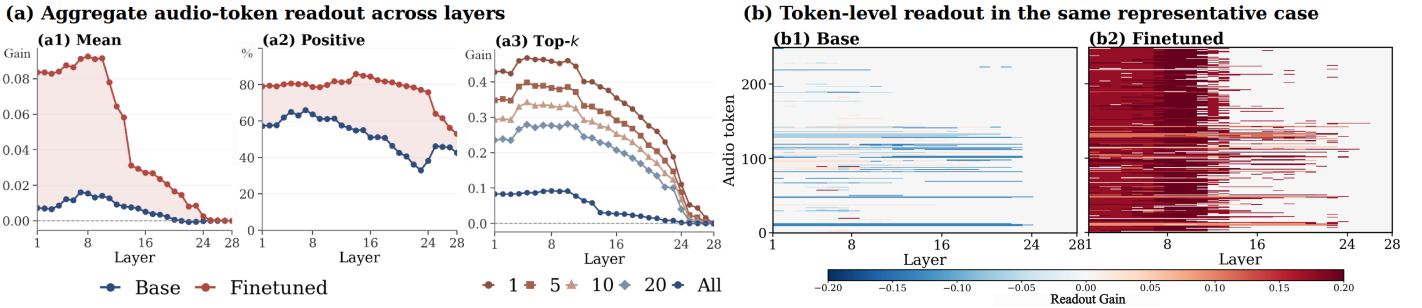}}
\caption{
Audio-token readout analysis.
(a1) Mean calibrated readout gain for base and fine-tuned models.
(a2) Percentage of tokens with positive gain.
(a3) Fine-tuned Top-\(k\) gain among highest-gain tokens.
(b) Token-level readout heatmaps for the representative case in Figure~\ref{fig:latentlens}.
}
\label{fig:patchscope}
\end{figure*}

\subsection{Semantic Scores and Predicted Support}

Figure~\ref{fig:latentlens} characterizes how temporal-grounding fine-tuning changes the correspondence between query-relevant semantics in audio-token space and the temporal support generated by the model. The analysis considers both the strongest event-aligned tokens and the broader distribution of semantic scores across layers.

Figures~\ref{fig:latentlens}(a1--a3) present the aggregate results. Figure~\ref{fig:latentlens}(a1) shows that fine-tuning consistently increases the fraction of Top-20 query-aligned tokens contained within the model-predicted temporal support, with larger differences emerging in later layers. Changes relative to annotated windows are weaker, indicating that the dominant effect is a closer correspondence between internal semantic evidence and the model's own temporal output. Figure~\ref{fig:latentlens}(a2) extends this observation to all audio tokens. Positive separation indicates that tokens inside the temporal support exhibit higher query-conditioned similarity than those outside it. Although the separation varies more across layers than the Top-20 hit ratio, it shows that the relationship between predicted support and event semantics is also reflected in the broader score distribution. Figure~\ref{fig:latentlens}(a3) summarizes these changes: predicted-window hit gains remain positive throughout the network and reach approximately $+11$--$13$ percentage points in the final layers, whereas annotated-window gains are weaker and often negative. Together with the checkpoint-stability analysis in Appendix~E.1, these results show that fine-tuning brings model-predicted temporal support into closer correspondence with largely stable query-relevant evidence already present in the audio-token stream.

Figure~\ref{fig:latentlens}(b) provides a token-level view for a representative example with multiple intervals. Early layers show limited differences across checkpoints, whereas middle and late layers exhibit clear recalibration of semantic scores while retaining much of the temporal structure of the tokens with the highest scores. At L28, the temporal distributions remain close, with $W_1(\mathrm{time})=0.2390$s, while the event similarity scores differ with $W_1(\mathrm{sim})=0.2938$. This case illustrates semantic-score recalibration over largely shared temporal support. Together, the aggregate and example-level results answer RQ1 by showing pre-existing event evidence at largely stable temporal locations and provide semantic evidence for RQ3 by showing closer alignment with predicted support after fine-tuning.

\subsection{Event Evidence Becomes More Accessible}
Figure~\ref{fig:patchscope} summarizes whether the event semantics identified in representation space are readable by the language decoder, using the calibrated readout gain from Section~\ref{sec:method_patchscope}.

Figures~\ref{fig:patchscope}(a1--a3) show that fine-tuning increases both mean readout gain and coverage of positive tokens, with the largest differences in early and middle layers. Mean readout gain attenuates toward zero in later layers, whereas coverage of positive tokens declines but remains substantial. The Top-$k$ analysis further shows that the largest gains remain concentrated among highly readable states, while the layerwise profile is stable across values of $k$. Fine-tuning therefore broadens decoder accessibility while concentrating its strongest effects in audio tokens with high responses.

Figure~\ref{fig:patchscope}(b) provides a token-level view for the same representative example analyzed in Figure~\ref{fig:latentlens}(b). Positive readout is sparse and fragmented in the base checkpoint, but becomes substantially denser after fine-tuning, particularly across early and middle layers. The increase spans a broad range of audio-token positions rather than a few isolated states. Together with Figure~\ref{fig:latentlens}(b), this case shows that event evidence organized over largely shared temporal locations becomes more broadly accessible to the language decoder after adaptation.

\begin{table}[t]
\centering
{\footnotesize
\setlength{\tabcolsep}{1.5pt}
\begin{tabular}{lcccc}
\toprule
Layers & Decoder swap & State swap & Gap & 95\% CI \\
\midrule
L1--L10  & $0.0929$ & $-0.0020$ & $0.0949$ & $\pm 0.0224$ \\
L11--L18 & $0.0496$ & $-0.0057$ & $0.0553$ & $\pm 0.0131$ \\
L19--L24 & $0.0192$ & $-0.0037$ & $0.0229$ & $\pm 0.0073$ \\
L25--L28 & $0.0003$ & $0.0000$  & $0.0003$ & $\pm 0.0004$ \\
\bottomrule
\end{tabular}
}
\caption{Cross-checkpoint readout control on 100 stratified instances. Values are layer-band mean readout-gain changes. Gap is decoder-swap minus patched-state-swap (95\% CI).}
\label{tab:readout_swap}
\end{table}

The bands coarse-grain the per-layer curves in Appendix~C into early, middle, late,
and near-output regions.
Although the audio tower and projector remain frozen, \(h_\ell^a(i)\) is produced by language-model blocks up to layer \(\ell\) and can therefore differ across checkpoints. We extract this layer-\(\ell\) audio-token hidden state from either checkpoint under the original audio-question context, patch it into the same layer of either checkpoint's diagnostic prompt, and let the remaining language-model layers complete the readout, yielding four patched-state and downstream-decoder combinations (Appendix~C).

Table~\ref{tab:readout_swap} reveals clear downstream-decoder dominance. Swapping the downstream decoder yields positive gains from L1--L24, whereas swapping the checkpoint that supplies the patched state has little effect; the gap confidence intervals exclude zero in the first three bands. These results answer RQ2: the early- and middle-layer gains in Figure~\ref{fig:patchscope} arise primarily from fine-tuned decoder access rather than from differences between base and fine-tuned intermediate states. Both effects approach zero in L25--L28, where little downstream computation remains after insertion.

\subsection{Output Consistency and Functional Relevance}

To connect diagnostic readout with timestamp generation, we test the pathway
inside the original generation context. On 100 fine-tuned Qwen2.5-Omni test
instances, the temporal-window probes from Section~\ref{sec:method_probe}
distinguish annotated-window recoverability from consistency with each
checkpoint's own predicted temporal support. Residual-delta erasure then
compares predicted-window audio-token masks with size-matched random
audio-token masks.

\begin{figure}[!t]
\centering
\includegraphics[width=\columnwidth]{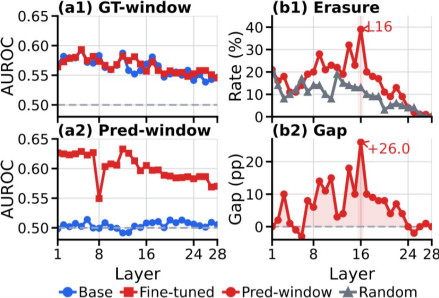}
\vspace{-1.0ex}
\caption{
Output-facing temporal diagnostics.
(a1,a2) Layerwise AUROC for membership in annotated and checkpoint-predicted
windows; dashed lines mark chance.
(b1,b2) Residual-delta erasure compares predicted-window and size-matched random
updates; (b2) reports the degradation gap.
}
\label{fig:output_diagnostics}
\end{figure}

Figures~\ref{fig:output_diagnostics}(a1,a2) reveal a complementary contrast
between annotated-window discrimination and predicted-window consistency. For
annotated windows, both checkpoints remain above chance and exhibit similar
layerwise profiles, indicating little change in linear recoverability after
fine-tuning. In contrast, the base predicted-window probe remains near chance
across layers, whereas the fine-tuned checkpoint achieves substantially higher
AUROC over most of the network. The position-only control in Appendix~E also
remains close to chance, showing that this separation is not explained by token
order alone. Thus, fine-tuning improves alignment with predicted support more
than recoverability of annotated event windows, answering RQ3.

Figures~\ref{fig:output_diagnostics}(b1,b2) show that erasing predicted-window
audio-token updates causes larger output degradation than erasing random
audio-token updates of the same size. At Layer 16, degradation is 39.0\% under
predicted-window erasure and 13.0\% under size-matched random erasure, yielding
a peak \(+26.0\) percentage-point gap (95\% CI: [15.0, 36.0]). Because masks are
size matched, this excess degradation answers RQ4 by showing that
predicted-window residual updates contribute more directly to timestamp
generation.

Together, these two views separate output consistency from generation-time
relevance: the probes measure alignment between states and emitted windows,
whereas residual-delta erasure tests whether updates at those windows affect the
timestamp answer.

\subsection{Cross Model Check}
For the non-intervention diagnostics, Appendix~G repeats the semantics, readout, and temporal-probe diagnostics on Qwen2-Audio-7B-Instruct. Predicted-window Top-20 hit gains are positive in all 28 layers, averaging 10.11 percentage points, predicted-window probes improve more clearly than annotated-window probes, and high-response readout strengthens after fine-tuning (Top-5 gain +0.0347, positive in 24/28 layers). Thus, improved readability and prediction alignment generalize, while layerwise concentration remains model dependent.

\section{Discussion}

\textbf{Latent Event Evidence (RQ1/RQ3).} The results suggest that grounding fine-tuning does not simply create event evidence from scratch. Fine-tuning substantially improves timestamp prediction, yet the base checkpoint already contains measurable event-related structure. The audio tokens most strongly aligned with the queried event appear at similar temporal positions before and after grounding fine-tuning, and annotated-window membership is weakly but consistently recoverable in the base checkpoint. Thus, poor grounding before fine-tuning is better explained by a weak mapping from existing event evidence to temporal outputs than by the absence of such evidence.

\textbf{Layered Semantics-to-Readout (RQ2/RQ3).} Adaptation appears to induce a layered division of labor rather than a single-layer bottleneck. Early and middle layers show the strongest gains in decoder accessibility: calibrated readout becomes stronger and more widespread, and the cross-checkpoint control attributes these gains primarily to decoder-side adaptation. Later layers show the closest correspondence between query-relevant semantic evidence and predicted temporal support, suggesting tighter coupling to output-facing timing decisions in later computation. Query-side activation patching in Appendix~F supports this progression: phrase-level influence appears earlier, whereas answer-facing start-time recovery emerges closer to the output. The Qwen2-Audio replication shows the same broad transition from latent evidence to improved readability and prediction alignment, although the layerwise concentration remains model dependent.

\textbf{Output Consistency and Relevance (RQ3/RQ4).} The results separate three increasingly stringent notions of temporal evidence: recoverability, output consistency, and generation-time relevance. The annotated-window probe measures whether temporal membership is recoverable from audio-token states, whereas the predicted-window probe measures consistency between those states and each checkpoint's own temporal outputs. Neither probe alone establishes that the recovered information affects generation. Residual-delta erasure addresses this gap: removing updates within predicted windows degrades timestamp generation more than removing random updates of the same size. This intervention does not identify a complete circuit, but it narrows the functional test to predicted-window residual updates during generation. The agreement among recoverability, output consistency, and the intervention during generation supports the conclusion that fine-tuning improves how existing event evidence is read out and coupled to output timing.

\section{Conclusion}
Temporal grounding provides a diagnostic setting for examining how language-model fine-tuning affects the semantics, decoder accessibility, and temporal output alignment of native audio-token states. In both LALMs studied, base models contain latent event evidence, while fine-tuning strengthens decoder readout and consistency with predicted windows. Residual-delta erasure further shows that predicted-window updates contribute to timestamp generation. Together, these findings support a semantics-to-readout account in which supervision improves access to existing event evidence and connects it more reliably to temporal outputs. Thus, temporal grounding acts as both an evaluation target and a compact probe of decoder-accessible audio evidence.

\bibliography{ref}

\clearpage
\includepdf[pages=-]{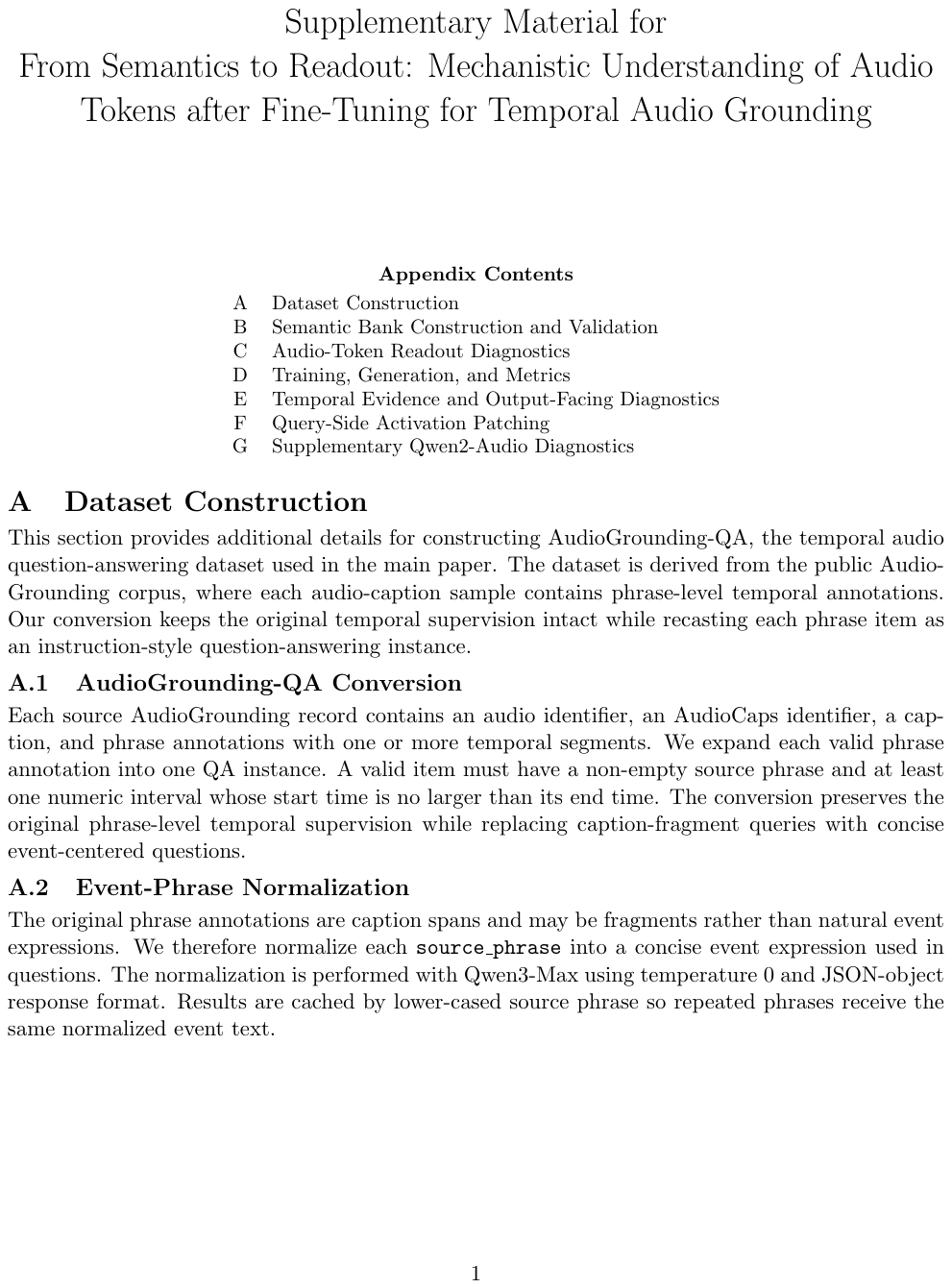}

\end{document}